\documentclass{PoS}

\usepackage{amsmath}
\usepackage{amssymb}
\usepackage{longtable}
\usepackage{xcolor}
\usepackage{color}
\usepackage{lipsum}
\usepackage{textcomp}

\usepackage{enumerate}

\usepackage{color}
\usepackage{tabulary}
\usepackage{multirow}
\usepackage{amsmath}
\usepackage{amsfonts}
\usepackage{ulem}

\usepackage{epsf}
\usepackage{epsfig}
\usepackage{pdflscape}
\usepackage{amsmath}
\usepackage{graphicx}

\usepackage{epsf,palatino,graphicx,wrapfig,array,setspace,amsmath,amssymb,fancyhdr,multirow,lscape,appendix,rotating,txfonts}
\usepackage{graphics,epsfig,txfonts}

\usepackage{graphicx}
\usepackage[flushleft]{threeparttable}
\usepackage{booktabs} 

\newcommand{\oergs}[1]{$10^{#1}$ erg s$^{-1}$}

\title{Intermediate Mass Black Holes: A brief review}

\ShortTitle{Review of IMBHs}

\author{\speaker{Filippos Koliopanos}\\
        CNRS, IRAP, 9 Av. colonel Roche, BP 44346, F-31028 Toulouse cedex 4, France \\
        Universit{\'e} de Toulouse; UPS-OMP; IRAP, Toulouse, France\\
        E-mail: \email{fkoliopanos@irap.omp.eu}}

\abstract{Intermediate mass black holes (IMBHs) are an (as yet) elusive class of black holes that are expected to lie in the $10^2-10^5\,M_{\odot}$ range, between the firmly established stellar-mass black holes and  ${\gtrsim}10^6\,M_{\odot}$ supermassive black holes. Predicted by a variety of theoretical models, IMBHs are the potential seeds of supermassive black holes and are expected to power some of the brightest extra-nuclear X-ray sources. This brief review is the result of a presentation and subsequent discussion of IMBHs that took place during the 12th International Frascati workshop on ``Multifrequency Behaviour of High Energy Cosmic Sources''. The manuscript aims to provide a concise and up-to-date review of the different evolutionary scenarios for the creation of IMBHs. Throughout the text I emphasize the importance of the identification and classification of IMBHSs in our effort to understand the formation of supermassive black holes and their co-evolution with their host galaxies.}

\FullConference{XII Multifrequency Behaviour of High Energy Cosmic Sources Workshop\\
		12-17 June, 2017\\
		Palermo, Italy}

\begin{document}

\section{Introduction}
First theorized by Schwarzschild (1916) and reformulated by Droste (1917) \cite{droste1917field}, Hilbert (1917) \cite{hil} and Weyl (1917) \cite{weyl} as peculiar repercussions of Einstein's theory of gravitation, black holes (BHs) were later predicted by Landau (1932) \cite{lan}, Chandrasekhar (1932, 1935) \cite{chan32}, \cite{1935MNRAS..95..207C} to be the end point of massive stars that collapse into singularities. However, it wasn't until 1972 and the dynamical estimation of mass of several $M_{\odot}$ for the ``dark'' companion of the Cyg X-1 binary star system, that an astrophysical source was firmly recognized as black hole (BH) \cite{1972Natur.235...37W}. Numerous subsequent observations of BH X-ray binaries (BH-XRBS: e.g.~\cite{2006ARA&A..44...49R,2014SSRv..183..223C}) have established the abundance of stellar mass BHs in the Universe, leading to the confirmation of BH mergers, in the recent discovery of gravitational waves \cite{2016PhRvL.116x1103A,2017PhRvL.118v1101A,2017PhRvL.119n1101A}.

In the 1970s Sanders and Lowinger (1972) \cite{1972AJ.....77..292S} used near infrared observations of the galactic center to infer the presence of a massive central object exceeding $10^5\,M_{\odot}$, while dynamical estimations suggested the presence of a supermassive black hole (SMBH) in the center of M87. Indeed Rees in 1984 \cite{1984ARA&A..22..471R} promoted the presence of SMBHs in the centers of massive galaxies. Approximately a decade later an abundant accumulation of solid observational evidence for the presence of SMBHs in active galactic nuclei (beginning with the remarkable detection of a relativistically broadened Fe K$\alpha$ line in MCG-6-30-15 \cite{1995Natur.375..659T}, H$_2$O maser emission from a $4{\times}10^7\,M_{\odot}$ SMBH in NGC~4258 \cite{1995Natur.373..127M}), dynamical estimations of the SMBH in the center of the Milky Way (e.g.~\cite{2002Natur.419..694S}), along with strong indications for weakly accreting SMBHs in most nearby galaxies, has convinced the vast majority of the astrophysics community that SMBHs lie in the centers of most -- if not all -- galaxies. 

The wealth of observational confirmations for the existence of smaller mass BHs (${\lesssim}200\,M_{\odot}$) and SMBHs (${\gtrsim}10^{6}\,M_{\odot}$) stands in stark contrast to the scarcity of observational evidence in favor of the existence of BHs between these two mass regimes. With the exception of very few strong candidates (see Section 4) and a modest sample of potential candidates, we have yet to confirm the existence intermediate mass black holes (IMBHs) -- which are defined as BHs with $100\,M_{\odot}{\lesssim}M_{\rm BH}{\lesssim}10^{6}\,M_{\odot}$ -- and are even further from establishing any characteristics of their population (i.e.~the existence of two sub-populations of 100-1000\,$\,M_{\odot}$ and $10^4-10^5\,M_{\odot}$). Establishing these facts is a crucial and necessary step in understanding stellar and galactic evolution as well as probing BH feedback in different mass regimes. 

In terms of stellar evolution, IMBHs are the expected relics of evolved Population III (Pop III) stars. Produced inside $10^6\,M_{\odot}$ dark matter mini-haloes at redshift z>15 (e.g.~\cite{1996ApJ...464..523H,2003ApJ...592..645Y}), Pop III stars can exceed 200\,$M_{\odot}$. Their high mass is the result of the lack of heavier elements in their progenitor cloud which maintains low fragmentation, as the gas cools inefficiently and the Jeans mass is higher. The resulting massive Pop III stars retain their mass as they exhibit diminished stellar winds than what is expected in massive metal rich stars (e.g.~\cite{1999ApJ...527L...5B,2002ApJ...564...23B,2001ApJ...548...19N,2002Sci...295...93A,2000ApJ...534..809O,2003ApJ...589..677O,2007ApJ...654...66O}). After their short lifetime (${\lesssim}3\,$Myr) Pop III stars may collapse into BHs that can exceed $200\,M_{\odot}$ (e.g.~\cite{2001ApJ...551L..27M,2004MNRAS.352..547R,2011ApJ...738..163W}). Discovery of IMBHs, in the $60-300\,M_{\odot}$ range, will provide tremendous insight into the Pop III stars that as of yet cannot be observed directly\footnote{Due to their large distance and low surface number densities Pop III stars will most likely be outside the reach of the upcoming James Webb Telescope \cite{2013MNRAS.429.3658R}}. Furthermore, IMBHs are also predicted to exist in accretion disks of active galactic nuclei (AGN, e.g.~\cite{2012MNRAS.425..460M,2014MNRAS.441..900M,2016ApJ...819L..17B}) and produced by repeated mergers of compact remnants of massive star in circumnuclear giant H II regions with a dense stellar population (e.g.~\cite{2000PASJ...52..533T} or onto other massive stars  in young massive clusters (e.g.\cite{2002MNRAS.330..232C,2002ApJ...576..899P,2004Natur.428..724P,2004ApJ...604..632G,2006MNRAS.368..141F}).

In terms of galaxy evolution, the detection -- and more importantly -- the determination of the spatial and mass distribution and the size of an IMBH population is a pivotal step in our understanding of formation of SMBHs and consequently galaxy formation itself. The discovery of luminous (L${\gtrsim}10^{47}\,$erg/sec) quasars at z${\sim}6$ reveals that SMBHs with $M_{\rm BH}{\gtrsim}10^{9}\,M_{\odot}$ must have formed within ${\lesssim}1$\,Gyr. This realization raises crucial questions with regard to the initial seeds of SMBHs and the process by which they are formed. The different formation paths will have a decisive impact on the star formation of the host galaxies. They can be divided into two main scenarios (that involve the presence of IMBHs) involving ``light'' or massive seeds. For a detailed review of the different formation paths of SMBHs see Volonteri et al.~2010  \cite{volonteri2010formation}.

In the light seed scenario, the SMBH forms either via (super-Eddington) accretion onto ${\sim}100\,M_{\odot}$ IMBHs (e.g.~\cite{Abel2002,Bromm2002,Turk2009,Tanaka}), or onto ${\sim}10^3\,M_{\odot}$ IMBHs seeds which in turn are the result of multiple mergers of Pop III IMBH primordial X-ray binaries (e.g.~\cite{2001ApJ...551L..27M,2003ApJ...593..661V,2016MNRAS.460.4122R}). In the massive seed -- or direct collapse -- scenario the SMBH seed is the result of direct collapse of primordial pristine hydrogen gas in Ly$\alpha$ cooling haloes with virial temperatures exceeding $10^4\,$K. In addition to the lack of metal line cooling, the halo requires the presence of strong Lyman-Werner UV radiation to destroy molecular hydrogen and prevent early collapse and fragmentation, until the structure collapses into an IMBH of ${\sim}10^4\,M_{\odot}$ (e.g.~\cite{2004ApJ...610...14W,2012ApJ...756L..19W,2012MNRAS.425.2854A,2017ApJ...842L...6W}). Depending on the mass of the seed different evolutionary paths lead to the creation of the SMBH, which may involve sustained Eddington limited accretion or episodic super- Eddington accretion (e.g.~Pacucci et al.~2015; Inayoshi et al.~2016). More importantly, the energy range of the resulting X-ray emission differs considerably, which in turn may result in positive or negative feedback on the formation of early galaxies. Hard X-rays (energies ${\gtrsim}$1\,keV) are more penetrative and may heat the intergalactic medium, suppressing star formation, while soft X-rays (0.1 - 1 keV: emitted from more massive accreting IMBHs accreting at lower rates) can promote the formation of H$_{2}$ by increasing the fraction of free electrons via photoionization, which in turn facilitates the cooling and collapse of gas, therefore promoting star-formation (e.g.~\cite{2001ApJ...553..499O,2001ApJ...560..580R,2003MNRAS.340..210G,2003MNRAS.338..273M,2005MNRAS.363.1069K,2007MNRAS.375.1269Z,2008MNRAS.387..158R,2008MNRAS.384.1080T} ).  Therefore, discovering and measuring the mass of a potential IMBHs population  is a major step in understanding stellar evolution and galaxy formation at different mass regimes. The  ``relics'' of the different formation paths are presented in an illustrative plot by Mezcua (2017)(\cite{2017IJMPD..2630021M}, Figure~1 in this article. See also Fig.1 in Greene 2012) in her recent review of IMBHs. In addition to the different ``sizes'' of IMBHs predicted by the two scenarios, the direct collapse scenario predicts a much lower occupation fraction in low-mass galaxies than the Pop III remnant scenario \cite{2008MNRAS.383.1079V} (but also see Shirakata et al.~2016 implications \cite{2016MNRAS.461.4389S}).

In addition to central SMBHs, primordial BH formation scenarios based on phase transitions in the early Universe (e.g.~\cite{1967SvA....10..602Z,1982PhRvD..26.2681H,1999PhRvD..59l4014J}) also predict the existence of IMBHs in galactic halos at a substantial distance from the galactic center \cite{2000hep.ph....5271R,2001JETP...92..921R,2005APh....23..265K}. While a considerable fraction of primordial IMBHs may merge to produce central SMBHs, it is likely that  a considerable fraction of them, will remain in the interemdiate mass regime forming a population of IMBHs in the galactic halos. Members of such a population may manifest themselves as hyper-luminous X-ray sources, if they accrete matter from a companion (e.g.~ESO~243-49~HLX-1: \cite{2009ApJ...705L.109G}, see relevant section on ultraluminous X-ray sources below).

Below, I will present a concise review of the different observational methods employed in the search for IMBHs and their current observational constraints. For an extensive, recent review on IMBHs, see Mezcua 2017~\cite{2017IJMPD..2630021M}.

 \begin{figure*}
       \resizebox{\hsize}{!}{\includegraphics[angle=0,clip,trim=0 0 0 0]{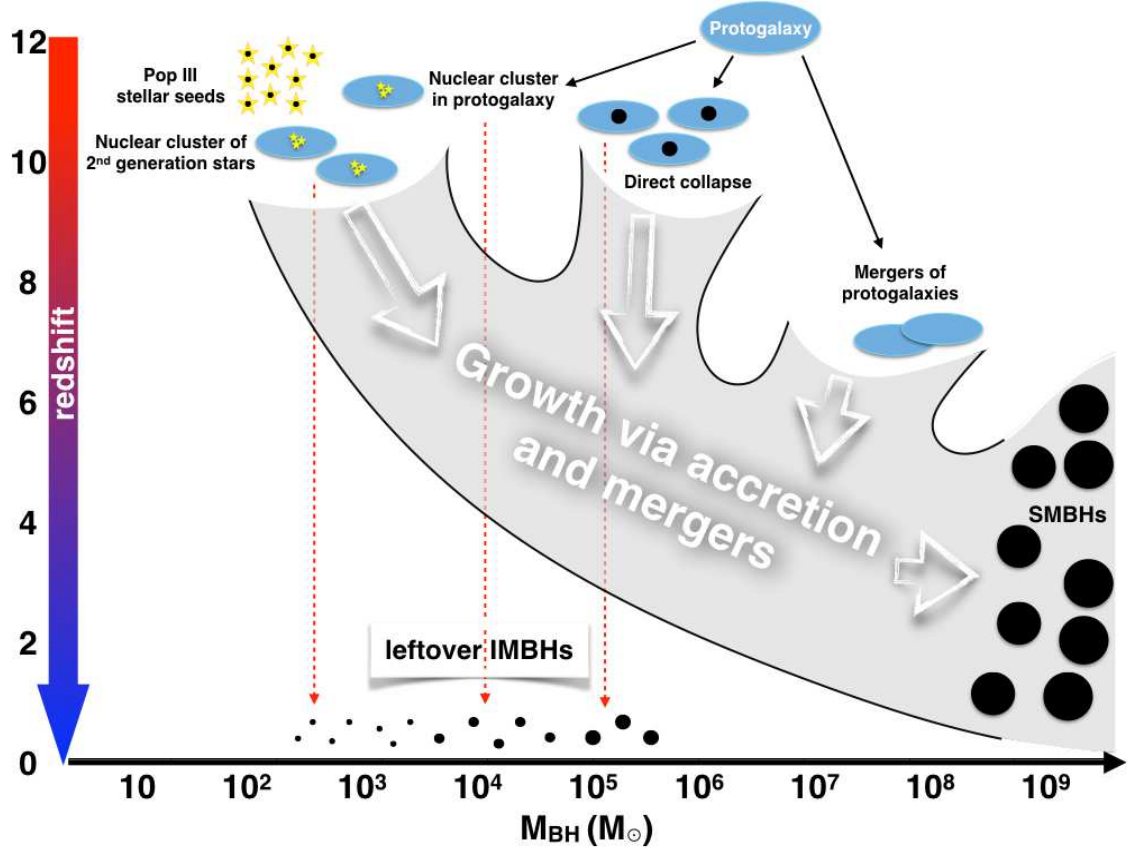}}
            \caption{ From Mezcua~(2017), with permission from the author and IJMPD: Formation scenarios for IMBHs. Seed BHs in the early Universe could form from Population III stars, from mergers in dense stellar clusters formed out either from the second generation of stars or from inflows in protogalaxies, or from direct collapse of dense gas in protogalaxies, and grow via accretion and merging to $10^9\,M_{\odot}$ by z${\sim}$7. SMBHs could also directly form by mergers of protogalaxies at z${\sim}$6. Those seed BHs that did not grow into SMBHs can be found in the local Universe as leftover IMBHs. }
   \label{fig:res}
 \end{figure*}

\section{Direct observational methods}

The most reliable method of detecting massive BHs (henceforth the term MBH will be used to refer to massive central BHs that may be either in the SMBH or the IMBH range) in the center of galaxies involves direct kinematic measurements. Such methods involve the dynamical measurements of individual sources, as is the case for Sgr A* (\cite{2002Natur.419..694S,2005ApJ...620..744G,2010RvMP...82.3121G,2012Sci...338...84M}), direct monitoring of gas motion orbiting the central MBH (e.g.~by measuring the offset of H$_2$O megamaser emission lines \cite{1995Natur.373..127M} or by 2D  modeling of the nuclear gas kinematics using  emission line imagery from high spatial resolution and/or adaptive optics assisted
observations (e.g.~\cite{macchetto1997supermassive, van1998evidence,kleijn2000black, barth2001evidence,sarzi2001supermassive,neumayer2007central}). 

The radius of influence of the BH can be expressed in terms of the velocity dispersion of the galactic spheroid ($\sigma_{\rm sph}$), namely $R_{inf}=GM_{\rm BH}/{\sigma_{\rm sph}}^{2}$ (e.g.~\cite{peebles1972star,frank1976effects}, $M_{\rm BH}$ is the BH mass). Inside the sphere of influence, objects are expected to follow Keplerian orbits under the gravitational force of the BH. For a BH mass of ${\sim}10^5\,M_{\odot}$ and $\sigma_{\rm sph}=30\,{\rm km/s}$ the radius of influence would be 0.5\,pc, making it impossible to resolve -- even for the nearest galaxies -- using current telescopes. As such, so far there are no definitive direct estimations of IMBHs, even for the nearest candidates. The nearby dwarf galaxy NGC 4395 (see Section 4) hosts one of the lowest directly measured BH masses, with a central MBH of $\log{M_{\rm BH}}=5-6\,M_{\odot}$, based on two-dimensional gas kinematic data \cite{den2015measuring}. This places the MBH in NGC 4395 at the lower end of SMBHs and tentatively in the IMBH range.  Dynamical gas modeling in NGC 404 yielded an estimation of ${\sim}10^5\,M_{\odot}$ \cite{seth2010ngc}, however this estimation is dependent on underlying model assumptions and is strained by uncertainties in the central stellar mass profile.  Only upper limits are provided for a handful more candidates (e.g.~\cite{valluri2005low,lora2009upper,jardel2012dark})

Another direct approach involves the reverberation mapping technique. Reverberation mapping involves accreting BHs and relies on the fact that the continuum emission of AGN is reprocessed in the high velocity photo-ionized gas that envelopes the central source -- known as the broad line region (BLR). Since the continuum emission is variable, time delays between fluctuations observed in the continuum and fluctuations of the velocity broadened emission lines of the BLR provides an estimate of the distance between the BH and the BLR, assuming light travel time. The Doppler broadening of the lines can be used to infer the velocity of the orbiting material in the BLR, yielding an estimation for the mass of the central BH assuming Keplerian rotation (for a detailed review of the method see e.g.~Peterson 2014 \cite{peterson2014measuring}). More precisely the value that is measured has dimensions of mass and is known as the virial product $M_{vir}=c{\tau}{\upsilon}^{2}/G$, where ${\tau}$ is the time lag and ${\upsilon}$ is the velocity of the material in the BLR. The actual mass of the BH is related to $M_{vir}$ by a factor of order unity, known as the virial factor (f), which is introduced by the kinematics and geometry of the BLR. 

In order to estimate the value of the virial factor, it is assumed that the $M_{\rm BH}$ values estimated by the reverberation mapping technique also follow the $M_{\rm BH}$-$\sigma_{\rm sph}$ distribution ( \cite{2000ApJ...539L...9F,2000ApJ...539L..13G}, see Section 3). The assumption is that for a given value of the $\sigma_{\rm sph}$  the $M_{\rm BH}$ estimated with the reverberation technique must be equal (within the relation's inherent scatter) with the value predicted using the dynamically estimated $M_{\rm BH}$-$\sigma_{\rm sph}$ relation and thus the virial factor is estimated for each source (e.g.~\cite{onken2004supermassive,collin2006systematic,greene2010precise,graham2011expanded,grier2013stellar}). Therefore any potential issues with the validity of the $M_{\rm BH}$-$\sigma_{\rm sph}$ results (e.g.~in the low-mass regime, see Section 3) will carry over to the virial mass estimations. Furthermore, recent analysis reveals that $M_{{\rm BH}}$ values derived from the accretion disk properties versus the virial mass estimates differ by a factor that is inversely proportional to the width of the broad emission lines, leading to significant viral mass mis-estimations \cite{2018NatAs...2...63M}. An example of the consequences of different modeling assumptions are evident in the mass estimation of the IMBH candidate in NGC 4395 using reverberation mapping. Edri et al.~(2012) measure a value of $M_{\rm BH}=(4.9{\pm}2.6){\times}10^4\,M_{\odot}$ \cite{edri2012broadband}, placing it decisively in the IMBH range, but Peterson et al.~(2005)  estimate  the mass to be much higher, at ${\sim}3.6{\times}10^5\,M_{\odot}$ \cite{peterson2005multiwavelength}.

\section{Indirect observational methods}

Direct observing methods are limited to only the most nearby galaxies and are vulnerable due to modeling assumptions. Furthermore, the reverberation mapping method requires multi epoch observations in order to measure the spectral time lag. In order to probe further and -- most importantly -- less luminous and less massive candidates, a variety of indirect methods can be employed. 

\subsection{IMBHs in AGN: Empirical relations between emission lines and/or optical continuum}

Virial mass estimates of central MBHs can be inferred using single epoch observations by making use of the empirical relation between optical AGN luminosity at 5100${\AA}$  and the size of the BLR, namely  $R_{BLR}{\sim}{L_{5100}}^{0.7}$ \cite{kaspi2000reverberation}. The virial mass can then be estimated by inferring the velocity of the BLR using the broadening of the H${\beta}$ emission line  (e.g., \cite{mclure2001black,mclure2002measuring,vestergaard2002determining,vestergaard2006determining,wang2009estimating}). Furthermore, by demonstrating a clear linear relation between the H$\alpha$ luminosity  and the optical continuum luminosity and a correlation between the widths of the H$\alpha$ and H$\beta$ lines, Greene \& Ho (2005) were able to infer the virial mass system based solely on observations of the broadened H$\alpha$ emission line. By way of optical spectroscopy and the use of classification techniques such as the ``Baldwin, Phillips \& Terlevich'' (BPT) diagrams~\cite{1981PASP...93....5B} (which distinguish between LINERs, H~II regions and AGN by measuring [O III]/H${\beta}$, [N II]/H${\alpha}$, and [S II]/H${\alpha}$ flux ratios), AGN are recognized among galactic surveys and their virial mass can be estimated by the above methods. This methodology has revealed more than 150 candidates with a central MBH in the $10^{5}\,M_{\odot}{\lesssim}M_{\rm BH}{\lesssim}10^{6}\,M_{\odot}$ range (e.g.~\cite{2004ApJ...610..722G,2013ApJ...775..116R}).

\subsection{Central IMBHs: Black hole mass - host bulge scaling relations}

Discovered simultaneously by two groups, Ferrarese \& Merritt 2000 and Gebhardt et al.~2000 \cite{2000ApJ...539L...9F,2000ApJ...539L..13G}, the scaling relation between the mass of the central MBH and the velocity dispersion of the stellar population of the spheroid bulge of galaxies ($M_{\rm BH}-\sigma_{\rm sph}$), demonstrated the high likelihood of a co-evolution between the central MBH and its host galaxy and sparked a keen interest in the study of different scaling relations between the $M_{\rm BH}$ and the properties of its host. The tight correlation between the mass of the central MBH and the velocity dispersion of the galactic bulge suggests that there is mechanical feedback between the MBH and it host galaxy, which extends well beyond the sphere of gravitational influence of the BH and transcends the merger and/or accretion history of the MBH. While the origin of the coupling between the MBH and the host bulge velocity dispersion is not yet resolved,
there is a multitude of analytic and numerical models that attempt to constrain the physical mechanism and reproduce the $M_{\rm BH}-\sigma_{\rm sph}$ relation (e.g \cite{silk1998quasars,haehnelt1998high,1999ASPC..182...87B,fabian1999obscured,haehnelt2000correlation,benson2003shapes,king2003black,wyithe2003self,granato2004physical,di2005energy,springel2005simulations,hopkins2005physical,sijacki2007unified,somerville2008semi,booth2009cosmological,zubovas2012m,costa2014feedback}). While their approach varies, all the above theoretical models rely on the presence of massive outflows or winds that are generated during the accretion episodes that are required to grow the central MBH. The models can be divided into  two broad categories depending on whether the interaction between the ouflow and the its host is ``energy-driven'' or ``momentum-driven''. The energy-driven models as suggested by Silk \& Rees (1998) \cite{silk1998quasars} reproduce the scale of the relation, i.e.~$M_{\rm BH}{\sim}{\sigma_{\rm sph}}^5$ as suggested by more recent observational campaigns by McConnell \& Ma 2013, but significantly underestimate the normalization of the relation. On the other hand, momentum-driven solutions as suggested by King et al.~(2003) \cite{king2003black} predict a slope of $M_{\rm BH}{\sim}{\sigma_{\rm sph}}^4$, but reproduce very well the $M_{\rm BH}$ for a given ${\sigma_{\rm sph}}$.

Nevertheless, the observationally constrained $M_{\rm BH}$-$\sigma_{\rm sph}$ relation itself is not well calibrated for low-mass galaxies, as the galactic samples employed to populate the relation suffer from luminosity bias and large uncertainties that are partially due to the precarious application of the $M_{{\rm BH}}$--$\sigma_{\rm sph}$ relation in the low-mass regime (e.g.~see discussion in \cite{2007ApJ...667..131G, 2014AJ....148..136M}). These issues result in an incompleteness of MBHs and host galaxies in the low-mass regime.

Another well-known scaling relation (and one which actually preceded the $M_{\rm BH}-\sigma_{\rm sph}$) is the one between the MBH mass and bulge luminosity ($M_{{\rm BH}}$--$L$: e.g.\cite{1989IAUS..134..217D, 1993nag..conf.....B,1998AJ....115.2285M,2003ApJ...589L..21M, 2013ApJ...764..151G,2016ApJ...817...21S}). 
The tight scaling relation further underlines the co-relation between SMBH and galaxy growth, posing questions as to how this is regulated. Namely, is the feedback from the accreting SMBHs that promotes the growth of its host galaxy, or is the environment created by stellar activity that enables the growth of the SMBHs? The $M_{{\rm BH}}$--$L$ relation can be used to infer the mass of the central MBH and has the advantage that it does not require  spectroscopy and can be applied to more distant galaxies.
Studying low-mass galaxies using these two relations can reveal new IMBH candidates (e.g.~\cite{2008AJ....136.1179B,2012ApJ...755..167D}). However, the sampling of these two relations is biased towards higher luminosities/masses, as most galactic samples with directly measured SMBH masses are dominated by luminous bulges (e.g.~\cite{2003ApJ...589L..21M,2007AAS...211.1327G}).  Furthermore, there are large uncertainties regarding the shape or even the application of the $M_{{\rm BH}}$--$\sigma_{\rm sph}$ relation in the low-mass regime (e.g see discussion in \cite{2007ApJ...667..131G, 2014AJ....148..136M,graham2016galaxy}). 

In recent studies, Graham (2012) \cite{2012ApJ...746..113G} and Graham \& Scott (2013) \cite{2013ApJ...764..151G} revisited the $M_{{\rm BH}}$--$L$ including many more low-mass spheroids with directly measured SMBH masses. Their work dramatically revised the relation, demonstrating that it most likely follows a broken power-law relation, becoming steeper at low- and intermediate-mass spheroids (see also \cite{2015ApJ...798...54G}. Using this improved scaling relation, \cite{2013ApJ...764..151G} discovered 40 low luminosity AGN (LLAGN) that  appear to have a mass of the central MBH that lies well within the IMBH range. However, Kormendy \& Ho (2013) \cite{kormendy2013coevolution} contest this result arguing that the steeper slope is the result of inclusion of pseudobulge galaxies, which are randomly offset towards lower masses and show little evidence of co-evolution between the central MBH and the host galaxy. On the other hand, Graham (2012,2015) argues that a (steeper than linear) relation does exist for pseudobulges (and low-mass classical bulges), which indeed appear to be offset towards lower masses (for similar values of $L$) \cite{graham2008populating,hu2008black}  and the ``broken'' $M_{{\rm BH}}$--$L$ holds. For a detailed discussion on this subject and also extensive reviews on SMBHs and their co-evolution with their host galaxies see Kormendy \& Ho (2013) \cite{kormendy2013coevolution} and Graham (2016) \cite{graham2016galaxy}.

Other MBH/host galaxy relations involve a correlation between the mass of a galaxy's central BH and its bulge concentration measured by the galactic S\'ersic index (e.g.~\cite{2001ApJ...563L..11G,2003RMxAC..17..196G,2007ApJ...655...77G,2013MNRAS.434..387S,2016ApJ...821...88S}). 
Furthermore, a correlation between the tightness of the spiral arms and the mass of the central MBH in disk galaxies was discovered by Seigar et al.~(2008) \cite{2008ApJ...678L..93S}. This correlation was formulated as a relation between the $M_{\rm BH}$ and the spiral pitch angle (PA) which measures the tightness of the spiral arms \cite{2008ApJ...678L..93S,2012ApJS..199...33D,2013ApJ...769..132B}. Each of these scaling relations may reveal possible candidates in the IMBH regime, however there are still large uncertainties in their estimations and the question of their validity in the low mass regime still remains. Nevertheless, in a recent publication \cite{koliopanos2017searching} four independent mass scaling relations and the fundamental plane of black hole activity (FP-BH: see next paragraph) were combined to measure the mass of the central MBH in seven LLAGN that had previously been reported to host IMBHs \cite{2013ApJ...764..151G}. While the study did not confirm the previous estimations (the MBHs had a $\log{M_{\rm BH}/M_{\odot}}{\approx}6.5$ on average), it demonstrated that all five (largely independent) methods produce consistent results in the low-mass regime. 
The study introduced this multiple-method/multi-wavelength approach as a go-to method for estimating central MBH masses when their spheres of gravitational influence cannot be spatially resolved. The combined techniques  protect  against outliers from any one relation and yield a more robust average prediction.

\subsection{IMBHs in AGN: The fundamental plane of black hole activity}

Another approach in the search for IMBHs draws on the well established relation between the X-ray emission, the radio emission and the mass of accreting BHs. These three quantities appear to be  strongly correlated, forming what is known as the ``fundamental plane of black hole activity'' (FP-BH \cite{2003MNRAS.345.1057M, 2004A&A...414..895F}).
During episodes of low-luminosity advection-dominated accretion (also known as a {\it hard state}), accretion disks are often accompanied by relativistic jets  (e.g.\cite{2003MNRAS.344...60G,2005Ap&SS.300..177N,2009MNRAS.396.1370F}). During this state of accretion the accretion disk flow is interrupted at some distance from the central object, at which point the material flows radially towards the BH. As a result of this new configuration, a ``corona'' of hot, optically-thin gas is created around the central object. When photons, from the (now truncated) accretion disk, traverse the ``corona'' they are Compton-upscattered to energies of tens to hundreds of keV. Furthermore the synchrotron  mechanism in the relativistic jets produces emission primarily at the radio band (e.g.~\cite{1977MNRAS.179..433B,1982MNRAS.199..883B,1984RvMP...56..255B}). Therefore the combined emission of the accretion disk, corona and relativistic jets ranges from the radio to the hard X-ray band (see Done 2007 \cite{2007A&ARv..15....1D} and Gilfanov 2010 \cite{2010LNP...794...17G} for reviews of accreting compact objects). The entire broadband emission can also be dominated by the jet itself, particularly if accretion rates are higher (${\sim}0.01\dot M_{\rm Edd}$, e.g.\cite{2004A&A...414..895F}).
An additional, less pronounced thermal component originating in the truncated accretion disk \cite{1973A&A....24..337S} may also be registered along with the primary non-thermal, coronal emission.
While the details of the disk-jet mechanism are still far from resolved it is understood that the jet is linked directly to the accretion process. More importantly, it has been shown that the luminosities of the radio and X-ray emission are correlated and the disk-jet mechanism is independent of the BH mass \cite{2003MNRAS.343L..59H}. Based on these theoretical predictions, Merloni et al.~(2003) \cite{2003MNRAS.345.1057M} and Falcke (2004) \cite{2004A&A...414..895F} probed a large sample of Galactic BHs and SMBHs and found a strong correlation between the radio luminosity ($ {\rm L_{r}}$, 5\,GHz), the X-ray luminosity ($ {\rm L_{x}}$, 0.5-10\,keV), and the BH mass, known as the FP-BH. 

The FP-BH can be used to infer the mass of accreting BHs during their hard-state. While this accretion state is more clearly defined in X-ray binaries (e.g.~\cite{2001ApJ...548L...9M, 2003MNRAS.345L..19M, 2006MNRAS.372.1366K}), the study of unobscured AGN shows that they can also be found in a similar accretion state (e.g.~\cite{1999MNRAS.304..160J,2000ApJ...530L..65N,2001MNRAS.327..739G,2002ApJ...564...86B,2009MNRAS.396.1929H,2011ApJ...733...60T,2012ApJ...745L..27P,2018MNRAS.474.1342M}) and therefore the method can be employed to put constraints on the mass of central MBHs. The most likely candidates to host IMBHs are LLAGN in low-mass galaxies as they  are expected to have undergone quiet merger histories and are, therefore, more likely to host lower-mass central BHs. While the X-ray and radio observations and the FP-BH can reveal IMBH candidates (e.g.~\cite{2014ApJ...787L..30R,koliopanos2017searching}), the scaling relation suffers from very high intrinsic scatter, therefore an estimation using solely the FP-BH would be very tentative. Nevertheless, as was demonstrated in Koliopanos et al.~\cite{koliopanos2017searching} it can provide a crucial, independent verification of estimations carried out using other methods.

\subsection{Extra-nuclear IMBHs}

\subsubsection{IMBHs and Globular clusters}

Under the right circumstances globular clusters (GCs) can become nurseries of IMBHs in the $10^3\,M_{\odot}$ range. Miller \& Hamilton (2002) \cite{coleman2002production} demonstrated that although consecutive three-body interactions may expel lighter BHs from GCs (e.g.~\cite{sigurdsson1993primordial,kulkarni1993stellar, 2018MNRAS.475.1574D}), more massive BHs (${\gtrsim}50\,M_{\odot}$) may sink towards the gravitational center of the cluster and after successive mergers with less massive BHs a  $10^3\,M_{\odot}$ IMBH is generated. Simulations have also shown that, in sufficiently dense GCs, stellar collisions could result in a stellar ``collision runaway'' during which repeated collisions create enormously massive stars which collapse into IMBHs (e.g.~\cite{zwart2002runaway,zwart2004formation,2004ApJ...604..632G,2006ApJ...649...91F}). While initial estimations of this process predicted IMBHs in the $10^3\,M_{\odot}$, further considerations of the uncertainties in the stellar evolution and mass loss of the merged systems showed that the resulting BHs could be as little as a few hundred $M_{\odot}$ mass \cite{2008A&A...477..223Y,2009A&A...497..255G, 2009Ap&SS.324..271V}. In the dense environment of GCs, IMBHs have a higher likelihood of capturing a companion. If the binding energy of the resulting binary is  greater than the mean stellar kinetic energy in the cluster (a so called ``hard binary''), repeated three body interactions will further harden the binary until accretion begins.

The presence of an IMBH -- in the $10^3\,M_{\odot}$ range -- in the center of a globular cluster is expected to effect its surrounding kinematics, in essence expanding the $M_{\rm BH}-\sigma_{\rm sph}$ relation to the lower mass regime. Estimations of the central velocity dispersion in GCs have revealed IMBH candidates in the GCs M15 and G1, (e.g.~\cite{bahcall1976star,peterson1989nonthermal, 1997AJ....113.1026G,2000AJ....119.1268G,2002ApJ...578L..41G,2005ApJ...634.1093G,2002AJ....124.3270G}). Integral field spectroscopy combined with surface brightness profiling and fitting to modeled slopes, predicted by N-body simulations, have also provided best fit estimations or upper limits for several more candidates (e.g.~\cite{2008ApJ...676.1008N,2010ApJ...719L..60N,2012A&A...542A.129L,2013A&A...552A..49L}). Nevertheless, there is considerable controversy over the interpretation of these measurements, as these results can also be explained by an increased central concentration of stellar BHs or neutron stars (e.g \cite{2003ApJ...582L..21B,2006ApJ...641..852V,2010ApJ...710.1032A,2010ApJ...710.1063V}). On the other hand, recent temporal studies of  millisecond pulsars in globular clusters have indicated the presence of a central IMBH. Perera~et al.~(2017)\cite{2017MNRAS.468.2114P} exploited 25 years of high precision observations of millisecond pulsar PSR B1820-30A, located in the globular cluster NGC 6624 to obtain rotational frequency time derivative measurements that can be interepreted as due to  orbital motion around a central IMBH with $M>7500\,M_{\odot}$. More recently K{\i}z{\i}ltan et al.~(2017) \cite{kiziltan2017intermediate} probed the dynamics of the central region of GC  47 Tucanae  by precisely measuring the location and acceleration of multiple millisecond X-ray pulsars found inside the cluster. Their results inferred the presence of a single, $2.3{\times}10^{3}\,M_{\odot}$ compact source (\cite{kiziltan2017intermediate}, see section 4).

Evidence for the existence of IMBHs in GCs and measurements of their mass can be obtained if the IMBHs are accreting matter from a companion, or from surrounding gas. Namely, X-ray and/or radio emission may reveal the presence of an accreting BH and its mass can be measured using the FP-BH. With the exception of GC G1, no X-ray emission or radio emission has been detected in those GC IMBH candidates that have their presence inferred from kinematic indications.  Radio observations of numerous GCs have so far only yielded  upper limits on the mass of potential IMBHs in their center (assuming that they are there and they are accreting, e.g.~\cite{2005MNRAS.356L..17M,2008MNRAS.389..379M,2008AJ....135..182B,2010MNRAS.406.1049C}, but have also raised questions on their existence or whether they accrete any matter (e.g.~Strader et al.~2010 \cite{2012ApJ...750L..27S} where no radio emission is detected in M15, M19, and M22, despite considerably deep observations or Wrobel et al.~2015 \cite{2015AJ....150..120W}, who detect no radio emission in more than 300 extragalactic GCs in NGC~1032 using stacked observations). Deep X-ray observations of $\omega$ Centauri for which kinematic estimations suggest the existence of an IMBH \cite{2008ApJ...676.1008N,2010ApJ...719L..60N} yielded no results, suggesting that either no IMBH is present or if it is, it is not accreting any mass \cite{haggard2013deep}. The only GC with detections of both X-ray and radio emission is G1 in galaxy M31 \cite{2006ApJ...644L..45P,2007ApJ...661L.151U,kong2010localization}, for which the FP-BH gives a consistent estimation with the $2{\times}10^4\,M_{\odot}$ reported by Gebhardt et al.~\cite{2002ApJ...578L..41G,2005ApJ...634.1093G}. However, we must note that Miller-Jones et al.~(2012)~\cite{2012ApJ...755L...1M} re-observed G1 M31 using the {\it VLA} telescope and detected no radio emission down to a 3${\sigma}$ upper limit of 4.7\,$\,mu$Jy.

\subsubsection{IMBHs and ultraluminous X-ray sources}
The search for IMBHs is not confined in the central MBHs of galaxies. Extra-nuclear IMBHs, 
produced inside dense globular clusters, can either sink towards to center of the cluster's gravitational potential well \cite{2004Natur.428..724P} or be  gravitationally ejected from the cluster \cite{holley2008gravitational}. Extra-nuclear IMBHs may also be the remnants of past interactions of their host galaxies with other dwarf galaxies. If an IMBH finds itself in a binary system and is accreting material from its companion, its mass may be inferred by limits imposed on its accretion rate by physical laws, i.e.~its Eddington limit. In accretion-powered sources the Eddington limit is imposed by the fact that when a certain threshold in the luminosity (corresponding to mass accretion rate) of the source is reached, the radiation pressure will exceed the ram pressure of the in-falling gas, thus hindering accretion and expelling the in-falling matter. In principle an accretion-powered source cannot be more luminous than its Eddington limit, which (as was shown by Eddington)  scales linearly with the mass of the central object. Assuming spherical accretion of hydrogen, the Eddington limit for a stellar mass BH is ${\sim}10^{39}$\,erg/s.

There are numerous off-nuclear, extragalactic X-ray sources with isotropic luminosities that exceed the Eddington limit for a stellar-mass BH, known as ultraluminous X-ray sources (ULXs) (see Kaaret 2017  \cite{2017ARA&A..55..303K} for a recent review). These sources were quickly considered as prime candidates for hosting IMBHs accreting at sub-Eddington rates in a similar fashion to standard stellar-mass XRBs (e.g.~\cite {1999ApJ...519...89C,2000ApJ...535..632M,2001MNRAS.321L..29K,2003ApJ...585L..37M}). While there is no known binary star evolution scenario that would result in an XRB with an IMBH accretor \cite{king2001ultraluminous}, an IMBH that has captured a companion could end up close enough with its companion to initiate mass accretion. This could either take place  inside the cluster or the two companions may be driven closely, due to multiple few-body dynamical recoils as the binary exits it \cite{gultekin2004growth,gultekin2006three,blecha2006close,o2006binary}. However such a scenario  would require most -- if not all ULXs -- to reside inside or close to globular clusters, which is not the case for the majority of ULXs. Furthermore, it has been demonstrated that most of the ULX population can be powered by stellar-mass BHs accreting at super-Eddington rates (e.g.~\cite{2003ApJ...596L.171G,2004NuPhS.132..369G,2004MNRAS.349.1193R,2007MNRAS.377.1187P,2009MNRAS.393L..41K}. The theoretical predictions are strongly backed by observations that reveal that the temporal and spectral characteristics of ULXs are markedly different from standard BH-XRBs, suggesting that they indeed are in a different state of (super-Eddington) accretion, dubbed the {\it ultraluminous state} \cite{2006MNRAS.368..397S,2009MNRAS.397.1836G,2013MNRAS.435.1758S,2015MNRAS.447.3243M}.
The astounding discoveries of a pulsating ULX \cite{2014Natur.514..202B,2017Sci...355..817I,2016ApJ...831L..14F} demonstrated that ULXs can also be powered by neutron stars. A number of compelling theoretical studies have suggested that it is very likely that many (or most) ULXs may indeed be powered by neutron stars rather than BHs (e.g.~\cite{2016MNRAS.458L..10K,2017MNRAS.468L..59K,2017MNRAS.tmp..143M}. Building on these theoretical predictions Koliopanos et al.~(2017) \cite{2017arXiv171004953K} demonstrated in a recently published study that the spectral characteristics of a large fraction of ULXs are consistent with accretion onto highly magnetized neutron stars rather than BHs.

Nevertheless, there is a small group of hyper-luminous X-ray sources (HLXs) with luminosities exceeding $\sim$\oergs{41} that are hard to explain even when invoking super-Eddington accretion and/or beaming of the X-ray emission. HLXs are strong IMBH candidates, with a particular source, known as ESO 243-49  HLX-1 (henceforth HLX-1), considered as the best IMBH candidate found so far (see details in next section). Indeed HLX-1 and also M82~X-1 -- the two brightest HLXs -- do not feature the spectral characteristics of ULXs (e.g.~\cite{2006ApJ...637L..21D,2009Natur.460...73F})
and seem to follow the transitions from soft to hard state similar to sub-Eddington stellar-mass BH-XRBs (ESO~243-49~HLX-1: \cite{2009ApJ...705L.109G,2011ApJ...743....6S,2012Sci...337..554W}, or M82~X-1: \cite{2010ApJ...712L.169F}).

\section{Notable candidates}

In this section, I present some notable IMBH candidates. The list is by far not comprehensive and only serves to present candidates with different origins, while also illustrating the results that can be achieved by different techniques.

\subsection{ULXs: HLX-1}

HLX-1 is probably the best IMBH candidate we know so far.  Discovered serendipitously by Farrell et al.~(2009) \cite{2009Natur.460...73F} in the 2XMM catalog \cite{2009A&A...493..339W}, HLX-1 had an unabsorbed X-ray luminosity of 1.1 $\times$ 10$^{42}$ erg s$^{-1}$ (0.2-10.0\,keV), for a distance of 95\,Mpc. Even if one assumes that the source is accreting at ten times the Eddington limit, this luminosity implies a minimum mass of 500 M$_\odot$ \cite{2009Natur.460...73F}, making HLX-1 a prime IMBH candidate. This claim relies on the assumption that HLX-1 is indeed located in galaxy ESO 243-49, which lies at the aforementioned distance of z=0.023. Indeed, in the years following the discovery of HLX-1 its association  with ESO 243-49 has been confirmed by numerous studies. Using the accurate X-ray position derived from {\it Chandra} data \cite{2010ApJ...712L.107W}, Soria et al.~(2010) \cite{2010MNRAS.405..870S} located the  source's optical counterpart, which shows an H$\alpha$ emission line with a similar redshift to that of ESO 243-49, confirming it as its host galaxy \cite{2010ApJ...721L.102W,soria2013kinematics}. The exact origin of the H$\alpha$ emission line is still unclear. If the line originates in an accretion disk \cite{lasota2011origin}, the disk must be observed almost face on, as the emission line is considerably narrow \cite{soria2013kinematics}. Wiersema et al.~(2010) \cite{wiersema2010redshift} suggested that the host star cluster or a photo-ionized/shock-ionized gas nebula very close to HLX-1 could be the source of the H$\alpha$ emission, while Soria et al.~2010 \cite{2010ApJ...721L.102W} suggest a low density ionized nebula as the source. A more recent study of observations taken with the MUSE instrument on the {\it VLT}, combined with complementary multi-wavelength data including {\it X-Shooter}, {\it HST}, {\it Swift}, {\it Chandra} and {\it ATCA} telescopes, revealed a strong variability of the emission line that was associated with the X-ray states of HLX-1. This finding confirms that the emission line originates close to the IMBH (negating the nebula or star cluster origin scenario) and validated the distance to HLX-1 \cite{2017A&A...602A.103W}.

HLX-1 -- similarly to standard BH-XRBs -- is also a source of occasional radio emission, which correlates with its transitions from hard (non thermal) to soft (thermal) spectral states \cite{2012Sci...337..554W}. Combining the X-ray and radio observations, the FP-BH allowed Webb et al.~(2012) to place an upper limit of less than $10^5\,M_{\odot}$ for the mass of the BH in the system. This estimation was further supported by the estimation of a mass of ${\sim}10^4\,M_{\odot}$ by Godet et al.~(2012) \cite{2012ApJ...752...34G}, Davis et al.~(2011) \cite{2011ApJ...734..111D}, and Straub et al.~(2014) \cite{2014A&A...569A.116S} using accretion disk modeling. The X-ray flux of HLX-1 varies by a factor of 50 as the source transitions between hard and soft states  \cite{2012ApJ...752...34G}. This transition, along with the periodical appearance of jet emission (as the most likely origin of the radio emission), is a crucial feature that sets HLX-1 out as an outstanding candidate for an accreting IMBH. ULXs -- which are most likely super-Eddington accretors --  do not transition between the strikingly distinct and characteristic ``hard'' and ``soft'' spectral states of standard  BH-XRBs ( e.g.~\cite{2008ApJ...687..471B,2010ApJ...724L.148G,2013MNRAS.435.1758S}) and most frequently feature a characteristic exponential spectral cutoff at ${\sim}10$\,keV, while the spectra of nominal BH-XRB extend to energies of often hundreds of keV during all spectral states. HLX-1, has a luminosity that is ten times higher than the brightest ULX and displays all the spectral and temporal characteristics of essentialy a scaled-up BH-XRBs, consistent with sub-Eddington accretion onto a very massive (${\gtrsim}$500\,$M{\odot}$ \cite{2009Natur.460...73F}) BH. 

The hyperluminous source \cite{2010ApJ...712L.169F} M82~X-1 has similar characteristics  to HLX-1. However, it  has not been observed as extensively as HLX-1, and recent works have shown that its spectrum can also be modeled assuming super-Eddington accretion onto a stellar mass BH \cite{2016ApJ...829...28B}. Other notable sources in the same category involve the ULX NGC~2276-3C which exhibits powerful jet emission both in the X-ray and radio wavelengths, with an FP-BH derived mass of ${\sim}5{\times}10^4\,M_{\odot}$ \cite{2013MNRAS.436.3128M,2015MNRAS.448.1893M} and the off nuclear X-ray source in NGC~5252 which radio and X-ray observations also place in the IMBH regime \cite{2015ApJ...814....8K}. More recent optical integral-field observations of the surrounding kinematics of the source suggest that it may be an off-nuclear AGN, belonging to a small low-mass galaxy that is being accreted by the Seyfert galaxy NGC~5252 \cite{2017ApJ...844L..21K}.

\subsection{Dwarf galaxies: NGC 4395}

NGC~4395 is an Sd type, bulgeless galaxy with a Seyfert 1 nucleus. The presence of an active MBH at the center of a bulgeless galaxy is exceptional in itself (see e.g.~Ferrarese et al.~2000 \cite{2000ApJ...539L..13G}, Merritt \& Ferrarese 2001 \cite{merritt2001black}), however the distinctness of NGC~4395 stems from the fact that it is one of the most likely candidates for a central, active IMBH. The presence of broadened optical and UV emission lines first revealed the AGN nature of NGC~4395 \cite{filippenko1993hst}. Compact radio emission \cite{ho2001radio}, suggestive of strong jet-like outflows \cite{wrobel2006radio} and highly variable and non-thermal X-ray emission (e.g.~\cite{moran1999nuclear,shih2003evidence,vaughan2005exceptional,moran2005extreme,nardini2011effects}) solidified its AGN nature. More remarkably, the source's central stellar velocity dispersion was less than ${\sim}$30\,km/s \cite{filippenko2003low}, indicating a mass of the central MBH that is less than $10^5\,M_{\odot}$ based on the $M_{\rm BH}$-$\sigma_{\rm sph}$ relation. This value is in agreement with estimations based on the broad profile of the H$\beta$ line and from X-ray variability properties \cite{filippenko2003low}. Further confirmation of a mass in the $10^4-10^5\,M_{\odot}$ range comes from reverberation mapping estimations (e.g.~\cite{peterson2005multiwavelength,edri2012broadband}) and from more recent kinematic estimations using integral field spectroscopy and fitting of the surface brightness of
the nuclear star cluster  \cite{den2015measuring}. Despite the detection of both X-ray and radio emission from the core of NGC~4395, estimating the mass of the MBH using the FP-BH has been proven difficult due to the source's very high variability and the fact that the X-ray flux variability is only weakly coupled to the radio flux. Assuming the mass estimated by Peterson \cite{peterson2005multiwavelength}, the analysis of contemporaneous {\it Swift} (X-ray) and Very Large Array (radio) observations has hinted at the possibility of NGC~4395 lying at a steeper track of the FP-BH \cite{king2013distinctive}. The presence of two distinct tracks in the FP-BH has been noted in stellar mass BH-XRBs, where state transitions occur at shorter intervals (e.g., \cite{corbel2003radio,corbel2012universal,gallo2003universal,gallo2012assessing}). In addition to NGC 4395, some other strong central IMBH candidates, in dwarf galaxies are: NGC~404, for which dynamical estimations provide an upper limit of $1.5{\times}10^5\,M_{\odot}$ for the mass of its central MBH \cite{nguyen2017improved}, an  estimate that is further supported by X-ray and radio observations of its active nucleus \cite{nyland2012intermediate,nguyen2017improved}; dwarf AGN POX~52, where multi-wavelength observations, including X-ray, radio, and H$\beta$ optical emission suggest the presence of a ${\sim}10^5\,M_{\odot}$ BH in its center (e.g.~\cite{thornton2008host}); LEDA 87000 (or RGG~118) whose central MBH lies in the intermediate mass regime, according to most BH mass - host bulge scaling relations \cite{2016ApJ...818..172G}, has the lowest BH mass measured in a dwarf galaxy and is also a low luminosity X-ray source \cite{baldassare201550}. Having had quiet merging and accretion histories, dwarf galaxies are strong candidates for hosting central IMBHs. A scenario that is further strengthened by recent extensive studies that indicate the presence of an IMBH population in dwarf galaxies up to a redshift of 2.4 \cite{2016ApJ...817...20M,2018arXiv180201567M}.

\subsection{Globular clusters: 47 Tucanae}

The discovery of an IMBH candidate in the center of globular cluster 47 Tucanae is compelling not just for its own sake, but also for proposing a novel method to infer the presence and measure the mass of a gas-starved IMBH that is not electromagnetically visible. Kinematic estimations of the mass of a potential IMBH in the center of 47 Tuc yielded inconclusive results as N-body simulations revealed that velocity dispersion measurements alone could not distinguish between predictions that did or did not include a central IMBH \cite{baumgardt2016n,mclaughlin2006hubble,watkins2015hubble}. K{\i}z{\i}ltan et al.~(2017)\cite{kiziltan2017intermediate} took advantage of the presence of 19 millisecond pulsars located in the cluster, for which high resolution timing analysis was available. The authors measured the pulsar acceleration, detecting additional components (to the intrinsic source spin-up), which they demonstrated can be attributed to the presence of a central MBH in the cluster. Using this method they inferred the mass of the IMBH candidate to a value of 2300$_{850}^{+1500}\,M_{\odot}$ and the mass of the globular cluster at 0.76${\times}10^6\,M_{\odot}$, in agreement with independent kinematic estimations \cite{baumgardt2016n,watkins2015hubble}. 

This novel method opens up the possibility for estimating the mass of other potential IMBHs in globular clusters for which electromagnetic detection is not feasible, but also to independently estimate the mass of IMBH candidates in GCs measured using more conventional methods.  Nevertheless, there are  caveats to the findings of K{\i}z{\i}ltan et al.. Ridolfi et al.~(2016)\cite{2016MNRAS.462.2918R} and Freire et al.~(2017)\cite{2017MNRAS.471..857F} obtained 23 timing solutions for pulsars in 47 Tuc, including spin period derivatives, but also orbital period derivatives, and proper motion and jerk measurements. The authors argued that  K{\i}z{\i}ltan et al.~had underestimated the distance to the cluster and by using an updated estimation for the distance,  Freire et al.~show that the all the accelerations of the pulsars in the clusters can be account for by a simple analytical model, without the need for a central IMBH.

\section{Conclusion: multi-wavelength future}

The extensive investigation by a large number of authors has yielded numerous tentative and several highly likely IMBH candidates, but a definitive detection of an IMBH has not happened yet. Nevertheless, the increasing number of plausible candidates discovered by multiple methods and at multiple wavelengths suggests that a population of IMBHs, very likely, exists but at this point its observational signatures lie at the threshold of our current observational limitations. 


Constraining this population and evaluating its characteristics is the next crucial step in the effort to understand how galaxies and globular clusters evolve. Determining the mass range in which central IMBHs lie (i.e.~in the $10^3-10^4\,M_{\odot}$ or ${\gtrsim}10^5\,M_{\odot}$ range) will largely decide on the formation path of SMBHs, as the different scenarios predict a starkly different shape of the $M_{\rm BH}$-$\sigma_{\rm sph}$ distribution in the low-mass regime (e.g.~Fig.~10 from Volonteri 2010 \cite{volonteri2010formation}). So far, galaxies with tentative measurements of their
central MBH and velocity dispersion (e.g.~\cite{greene2007new,ho2016low}) suggest the presence of a plume-like concentration of sources at the $10^5\,M_{\odot}$ mark (see also Volonteri 2010 \cite{volonteri2010formation} Mezcua 2017 \cite{2017IJMPD..2630021M} and references therein). While this preliminary, rough finding hints towards the dominance of the massive seeds scenarios, it is equally plausible that the lower mass IMBHs remain below our observational capabilities as their AGN are luminosity and distance limited and the $M_{\rm BH}$-$\sigma_{\rm sph}$ distribution is populated primarily by MBHs that have ended up above the IMBH regime, via accretion and mergers.
It becomes apparent from the multitude of studies, briefly reviewed above, that multi-wavelength observations hold the key to constraining the missing population of IMBHs. As we enter the era of multi-messenger astronomy and push the threshold of our observational capabilities, the detection and classification of IMBHs promises to be one of the big discoveries of the next decade.

\bigskip
\bigskip
\noindent {\bf DISCUSSION}

\bigskip
\noindent {\bf DANIELE FARGION:} I would like to also point out the case of M82 X-1 as well known candidate for an IMBH [Author's note: M82 X-1 had not been discussed during the presentation ].

\bigskip
\noindent {\bf FILIPPOS KOLIOPANOS:} Indeed, M82 X-1 is one of the two best candidate IMBH  hyperluminous sources. The source shows similar characteristics as HLX-1 and appears to behave like a sub-Eddington BH-XRB in the intermediate mass range. The reason the source was not mentioned in the oral presentation and is not discussed as extensively as HLX-1 in the review, is that M82 X-1 has not been monitored as extensively as HLX-1 which has been systematically observed at all wavelength, and also the fact that more recent estimations point out that M82 X-1 can be plausibly described as a super-Eddington source.

\bigskip
\noindent {\bf WOLFGANG KUNDT:} How do you explain the problem of the naked singularity?
[Author's note: Prof. Kundt has argued for a long time against the existence of BHs, and has also criticized Roger Penrose's cosmic censorship hypothesis. See e.g.~{\it Astrophysics Without Black Holes, and Without Extragalactic Gamma-Ray Bursts} \\ 
https://ojs.cvut.cz/ojs/index.php/APP/article/view/APP.2014.01.0027 ]

\bigskip
\noindent {\bf FILIPPOS KOLIOPANOS:} As I am not a theoretical physicist or mathematician, I cannot comment -- at an expert level -- on the potential problems of the cosmic censorship hypothesis. However, as an observer I am satisfied by the currently available observational evidence in favor of the presence of SMBHs -- and more importantly their interpretation in the context of general relativity. By ``evidence'' I am mostly referring to three observational milestones in the SMBH hypothesis. The detection of highly broadened, 6.4\,keV iron K$\alpha$ lines in AGN  (starting with the galaxy MCG-6-30-15 by Tanaka et al.~1995), that are consistent with gravitational redshift by compact objects with a mass that exceeds $10^6\,M_{\odot}$. The independent estimations from megamaser emission (e.g.~Miyoshi et al.~1995) displaying clear Keplerian rotation curves that require a mass of the order of ${\sim}10^{7}\,M_{\odot}$ and finally the dynamical evidence for the SMBH in the center of the Milky Way (e.g.~Sch{\"o}del et al.~2002).

\section{Acknowledgments}
The author would like to thank Mar Mezcua for refereeing and significantly improving this review and Franco Giovannelli for giving me the opportunity to present this work and be part of the XII Multifrequency Behaviour of High Energy Cosmic Sources Workshop .

\bibliographystyle{JHEP}
\bibliography{/home/filippos/Documents/Bibliography/general_scholar}

\end{document}